\documentstyle[preprint,aps]{revtex}
\begin{document}
\preprint{OITS-533, UH-511-779-94}
\draft
\title{Gluon dipole penguin contributions to $\epsilon'/\epsilon$
and CP violation in Hyperon decays in the Standard Model}
\author{N.G. Deshpande$^1$, Xiao-Gang He$^1$ and S. Pakvasa$^2$}
\address{$^1$Institute of Theoretical Science\\
University of Oregon\\
Eugene, OR97403-5203, USA\\
and\\
$^2$ Department of Physics and Astronomy\\
University of Hawaii\\
Honolulu, Hawaii, HI 96822, USA}
\date{Jan.. 1994}
\maketitle
\begin{abstract}
We consider the gluon dipole penguin operator contributions
to $\epsilon'/\epsilon$ and CP violation in hyperon decays. It has been
proposed by Bertolini et al. that the contribution to $\epsilon'/\epsilon$
may be significant. We show that there is a cancellation in the leading order
contribution and this contribution is actually suppressed by
a factor of order O($m_\pi^2,
m_K^2)/\Lambda^2$. We find
that the same operator also contributes to CP violation in hyperon decays where
it is not suppressed. The
 gluon dipole penguin operator can enhance CP violation in hyperon decays by
as much as 25\%.
\end{abstract}
\newpage
In this paper we study the gluon dipole penguin operator $\bar s
\sigma^{\mu\nu}t^aG^a_{\mu\nu} (1 - \gamma_5) d$ contributions to
$\epsilon'/\epsilon$ and CP violation in hyperon decays in the Standard Model
(SM). The effective $\Delta S = 1$ Hamiltonian at leading order
can be
parametrized as
\begin{eqnarray}
H_{eff} = {G_F\over \sqrt{2}} V_{ud}V_{us}^*\sum_iC_i(\mu)Q_i(\mu)\;,
\end{eqnarray}
where the sum is over the effective operators $i = 1 - 10$ defined in
Ref.\cite{buras}, and
the operators
\begin{eqnarray}
Q_{11} = {g_s\over 16\pi^2} m_s \bar s \sigma_{\mu\nu} t^a G^{\mu\nu}_a
(1-\gamma_5)d\;,\nonumber\\
Q_{12} = {eQ_d\over 16\pi^2} m_s \bar s \sigma^{\mu\nu}F_{\mu\nu}
(1-\gamma_5) d\;,
\end{eqnarray}
where $G^{\mu\nu}_a$ and $F^{\mu\nu}$ are the gluon and photon field strengths,
respectively. $t^a$ is the $SU(3)_C$ generators normalized as $Tr(t^at^b) =
\delta^{ab}/2$. $C_i = z_i + y_i\tau$ with $\tau =
-V_{td}V_{ts}^*/V_{ud}V^*_{us}$. CP violation is proportional to $y_i$.
The QCD corrected coefficients $y_i$
in the SM
have been evaluated in Ref.\cite{buras} and Ref.\cite{bert}.
In our later calculation we will use the values in Ref.\cite{bert} for $y_i$.
The contributions to
$\epsilon'/\epsilon$\cite{buras} and CP violation in hyperon
decays\cite{dp,dhp,hsv} from the operators $Q_1 - Q_{10}$ have
been extensively studied before. The contributions to $\epsilon'/\epsilon$
from $Q_{11,12}$ have been considered recently by Bertolini et al.\cite{bert}.
The dominant contributions come from the internal top quark. It was claimed
that
the contribution from $Q_{11}$ is sizable, while the contribution from
$Q_{12}$ is
negligible. In this paper we reconsider the $Q_{11}$ contribution to
$\epsilon'/\epsilon$ and also consider the contribution to CP violation in
hyperon decays. We find that there is a cancellation in the
contribution to $\epsilon'/\epsilon$ and the result obtained in Ref.\cite{bert}
is strongly
suppressed. There are substantial contributions to CP violation in some of the
hyperon decays.

\noindent
{\bf Contribution to $\epsilon'/\epsilon$}

The parameter $\epsilon'/\epsilon$ is a measure of CP violation in $K_{L,S}
\rightarrow 2 \pi$ decays. It is defined as
\begin{eqnarray}
{\epsilon'\over \epsilon} = i{e^{i(\delta_2-\delta_0)}\over
\sqrt{2}(i\xi_0 +\bar \epsilon)}\left |A_2\over A_0\right |\left (
\xi_2 - \xi_0\right )\;,
\end{eqnarray}
where $\bar \epsilon\approx 2.27\times 10^{-3} e^{i\pi/4}$ is the CP violating
parameter in $K^0 - \bar K^0$,
$\delta_i$ are the strong rescattering phases,
$ |ReA_2/ReA_0| \approx 1/22$, and $\xi_i = ImA_i/ReA_i$. Here  $A_0$ and $A_2$
are the decay amplitudes with $ I = 0$ and $2$ in the final states,
respectively.
In order to evaluate the $Q_{11}$ contribution to $\epsilon'/\epsilon$, one
needs
to calculate the matrix element $<\pi^0 \pi^0| Q_{11}|\bar K^0>$. $Q_{11}$ is
a $\Delta I =1/2$ operator which contributes to $ImA_0$ only.
If one uses the
 naive PCAC result as in Ref.\cite{bert}, one obtains
\begin{eqnarray}
<\pi^+ \pi^-|Q_{11}|\bar K^0> = {g_sm_s\over 16\pi^2 f_\pi^2f_K}
\left [<0|\bar d \sigma_{\mu\nu}t^aG^{\mu\nu}_a d|0> + <0|\bar s
\sigma_{\mu\nu}
t^aG^{\mu\nu}_a s|0>\right ]\;,
\end{eqnarray}
where $f_\pi = 132 MeV$ and $f_K = 161 MeV$. The matrix element on the
right-hand side of eq.(3) can be related to the quark condensation by
\begin{eqnarray}
g_s<0|\bar q \sigma_{\mu\nu} t^a G_a^{\mu\nu} q|0>
= m_0^2 <0|\bar q q|0>\;,
\end{eqnarray}
where $m_0^2 \approx 1\;GeV^2$\cite{mo} is a phenomenological constant.
Using
\begin{eqnarray}
<0| \bar u u + \bar s s|0> = -{f_K^2 m_K^2\over m_u + m_s}\;,
\end{eqnarray}
and  $<0|\bar d d |0> = <0|\bar u u|0>$, one finally
obtains
\begin{eqnarray}
<\pi^+ \pi^- |Q_{11}| \bar K^0> = -{m^2_0 m_s\over 16\pi^2 (m_u +m_s)}
{m^2_K f_K\over f^2_\pi}\;.
\end{eqnarray}
One also has,
$<\pi^0\pi^0|Q_{11}|\bar K^0>=<\pi^+ \pi^- |Q_{11}| \bar K^0>$.

Using the matrix element in eq.(7),  a quite large
contribution to $\epsilon'/\epsilon$ was obtained in Ref.\cite{bert}.
They find that $Q_{11}$ contributes between $(2 \sim 3)\times 10^{-4}$.
It weakly depends on the top quark mass $m_t$ for $m_t$ between $100\; GeV$
to $250\; GeV$.
We would like to point out that the
naive PCAC result obtained above is not correct. The calculation in
Ref.\cite{bert} is only part (Fig. 1.a) of the contributions.
An important  "tadpole" contribution (Fig. 1.b) was not considered in the
analysis of Ref.\cite{bert}.
This
contribution cancels exactly the PCAC result obtained above. The
net contribution is much smaller. The situation is the same as for
$\epsilon'/\epsilon$ in the Weinberg model of CP violation\cite{dh,hyc}. The
importance of the "tadpole" contribution was first noticed by Donoghue and
Holstein\cite{dh} in the Weinberg model of CP violation.
We now calculate these contributions in the SM.

The leading order chiral realization of $Q_{11}$ for $\bar K^0
\rightarrow n\pi$ is
\begin{eqnarray}
L = aTr(\lambda_6 +i\lambda_7)U) + H.C.\;.
\end{eqnarray}
The conventional parametrization of U is given by
$U = exp(-i4t^a\phi^a/f_\pi)$ with $\phi^a$ being the fields of
the pseudoscalars. In our discussion we follow Ref.\cite{hyc} using a
more general parametrization of U. To fourth power in the meson fields,
there is a free parameter $a_3$\cite{hyc,cor},
\begin{eqnarray}
U = 1 + i{2\over f_\pi}(2t^a\phi^a) - {2\over f_\pi^2}(2t^a\phi^a)^2
-i{a_3\over f_\pi^3} (2t^a\phi^a)^3 + {2(a_3-1)\over f_\pi^4}(2t^a\phi^a)^4 +
...
\end{eqnarray}
The conventional parametrization corresponds to $a_3 = 4/3$.
The on-shell amplitudes should not depend on $a_3$.
The effective lagrangian in eq.(8) will not only
generate the direct $K \rightarrow 2\pi$ amplitude (Fig. 1.a), but
 also generate a non-zero
$K\rightarrow vaccum$ transition amplitude. This non-zero $K\rightarrow
vaccum$ transition amplitude,
when combined with the strong $K\pi \bar K\pi$ amplitude
$A_{strong}(K\bar K\pi^0\pi^0)$ with one $K$ off-shell with $q^2 = 0$, will
also generate
a $K \rightarrow 2\pi$ amplitude as show in Fig. 1.b. The total amplitude
for $K\rightarrow 2 \pi$ from $Q_{11}$
is the sum of contributions from Fig. 1.a and Fig. 1.b.
We have
\begin{eqnarray}
&A&_{total}(\bar K^0 \rightarrow \pi^0 \pi^0) = A(fig. 1.a) + A(fig.
1.b)\nonumber\\
&=&A(\bar K^0 \rightarrow \pi^0\pi^0)|_{fig.1.a} + A_{strong}(K^0\bar K^0
\pi^0\pi^0)
 {1\over m_K^2} A(\bar K^0 \rightarrow vaccum)\;,
\end{eqnarray}
where $A(\bar K^0 \rightarrow n\pi) = <n\pi|-H_{eff}(Q_{11})|\bar K^0>$, and
$H_{eff}(Q_{11}) = -i{G_F\over \sqrt{2}}y_{11}Im(V_{td}V^*_{ts})Q_{11}$. We
have
\begin{eqnarray}
&A&( \bar K^0 \rightarrow vaccum) = -i
{G_F\over \sqrt{2}}y_{11}Im(V_{td}V^*_{ts})f_\pi\sqrt{2}
{g_sm_s\over 32\pi^2}\tilde A_{\bar K\pi}\;,\nonumber\\
&A&(\bar K^0 \rightarrow \pi^0) = -{G_F\over
\sqrt{2}}y_{11}Im(V_{td}V^*_{ts}){g_sm_s\over 32\pi^2}\tilde A_{\bar
K\pi}\;,\nonumber\\
&A&( \bar K^0 \rightarrow \pi^0 \pi^0) = i{G_F\over
\sqrt{2}}y_{11}Im(V_{td}V^*_{ts}){g_sm_s\over  32\sqrt{2}\pi^2f_\pi}\tilde
A_{K\pi}{a_3\over 2}\;.
\end{eqnarray}
Here $\tilde A_{K\pi} = -i<\pi^0|\bar s \sigma^{\mu\nu}2t^aG^a_{\mu\nu}
(1-\gamma_5) d|\bar K^0>$. Note that $A(\bar K^0\rightarrow \pi^0\pi^0)$ is
$a_3$
dependent, which can not be the final answer. Additional contribution from
Fig. 1.b has to be considered. What has
been evaluated in Ref.\cite{bert}  conrresponds to $A(fig. 1.a)$
with $a_3 =2$. $m_s\tilde A_{K\pi}$
has been calculated  to be
$0.11 \sim 0.17\; GeV^4$ in the MIT bag model\cite{dh1}. Using this value one
obtains approximately
the same numerical
value for $A(fig.a)$ as obtained in Ref.\cite{bert}.

To calculate the contribution from Fig. 1.b, one needs
 to know the strong $K\pi \bar K\pi$ amplitude $A_{strong}$. This can be
obtained
from the leading chiral largrangian
\begin{eqnarray}
L = {f_\pi^2\over 8} [Tr\partial_\mu U \partial ^\mu U +
B Tr(MU + U^\dagger M)]\;,
\end{eqnarray}
where $M = Diag (m_u\;, m_d\;,m_s)$ is the quark mass matrix and $B$ is a
constant. From this we obtain\cite{hyc,cor}
\begin{eqnarray}
 A_{strong}(K^0\bar K^0 \pi^0\pi^0) = {m^2_K\over 2f_\pi^2}{a_3\over 2}\;.
\end{eqnarray}
Here we have set the momentum of $ K$ annihilated into the vacuum
to be zero and others to be
on-shell (Of course the on-shell $A_{stong}$ is $a_3$ independent.).
Inserting eq.(13) into eq.(10), we find the $a_3$ dependent
contribution from Fig. 1.b
cancels exactly
that from Fig. 1.a. To this order, there is no contribution to
$\epsilon'/\epsilon$ from $Q_{11}$. However, the cancellation may not be
complete due to higher derivative terms in chiral perturbation theory.
Unfortunately, one does not know how to calculate higher-order
contributions to the
matrix elements. One can define a suppression factor D,
\begin{eqnarray}
A_{total}(K^0 \rightarrow \pi^0\pi^0) = A(fig.1.a) D\;.
\end{eqnarray}
The suppression factor should be of order $O(p^2)/\Lambda^2 \approx
O(m_K^2,m_\pi^2)/\Lambda^2$\cite{dh,hyc}. Here $\Lambda \approx 1\; GeV$ is
the chiral
symmetry breaking scale. The contribution is suppressed by a factor of 0.3 or
even more. The contribution from $Q_{11}$ calculated in Ref.\cite{bert}
correspond to $D = 1$.

We list the results for $D = m_K^2/\Lambda^2$ in Table 1, but note that
the sign of the contribution is unknown.  The parameter $Im(V_{td}V^*_{ts})$
is constrained
by the parameter $\epsilon$ from $K^0 -\bar K^0$ mixing. Unfortunately it is
not completely fixed. It can vary between $3 \times 10^{-4}$ to $10^{-4}$
for top quark mass $m_t$ about $100\;GeV$ and $2\times 10^{-4}$ to $0.5 \times
10^{-4}$ for $m_t$ about $200\;GeV$\cite{CP}. In Table 1, following
Ref.\cite{bert}, we use an approximation $Im(V_{td}V_{ts}^*) =
2.77\times 10^{-4}(m_t^2/m_W^2)^{-0.365}$ as the
central value. We see that the effect of $Q_{11}$ is generally insignificant,
but becomes important when the contribution from $Q_1 - Q_{10}$ become small
due to cancellations for $m_t$ around $200\; GeV$.
\newpage
\noindent{\bf CP violation in hyperon decays}

CP violation in hyperon decays in the SM has been studied
before\cite{dp,dhp,hsv}. The $Q_{11}$
contributions were not included, and we now turn to study these contributions.
Non-leptonic hyperon decays proceed into both S-wave (parity-violating) and
P-wave (parity-conserving) final states with amplitudes S and P, respectively.
One can write the amplitude in the rest frame of the initial baryon as
\begin{eqnarray}
Amp(B_i \rightarrow B_f\pi) =S + P\vec \sigma\cdot \vec q\;,
\end{eqnarray}
where $\vec q$ is the momentum of pion. It is convenient to write the
amplitudes
as
\begin{eqnarray}
S &=& \sum_i S_i e^{i(\phi^S_i + \delta_i^S)}\nonumber\\
P &=& \sum_i P_i e^{i(\phi^P_i + \delta_i^P)}\;
\end{eqnarray}
to explicitly separate the strong rescatering phases $\delta_i$ and the weak CP
violating phases $\phi_i$. In the rest frame of the initial baryon,
one particularly interesting observable
is the asymmetry A defined in Ref.\cite{dhp}
\begin{eqnarray}
A = {\alpha+\bar \alpha\over \alpha-\bar \alpha}\;,
\end{eqnarray}
where $\alpha = 2 Re(S^*P)/(|S|^2 +|P|^2)$, and $\bar \alpha$ is the
corresponding quantity for anti-hyperon decays. A non-zero A signals CP
violation. In this paper we will concentrate on the study of A.
The results can be easily generalized to other CP violating observables
defined in Ref.\cite{dhp}.
We will calculate the CP violating observable A  for $\Lambda
\rightarrow N \pi$ and $\Xi \rightarrow \Lambda \pi$. Since to leading
order the observables
$A(\Lambda^0_-)$ and $A(\Xi^-_-)$
for $\Lambda \rightarrow p\pi^-$ and $\Xi^- \rightarrow \Lambda \pi^-$ are
the same as $A(\Lambda^0_0)$ and $A(\Xi^0_0)$ for
$\Lambda \rightarrow n\pi^0$ and $\Xi^0 \rightarrow \Lambda \pi^0$,
respectively,  we choose to work with $\Lambda \rightarrow n\pi^0$ and $\Xi^0
\rightarrow \Lambda \pi^0$ decays.

For
the same reason as for the $K\rightarrow 2\pi$ amplitude, when evaluating the
S-wave amplitude, one should also include the contributions from the
direct contribution of Fig. 2.a and the "tadpole" contribution of Fig. 2.b. We
have\cite{dhp,dh2}
\begin{eqnarray}
S(\Lambda\rightarrow n\pi^0) &=&-{i\over \sqrt{2} f_\pi} <n
|H^\dagger_{eff}(Q_{11})|\Lambda>|_{fig.2.a} \nonumber\\
&+& {i\over \sqrt{2}f_\pi}\left [{3\over 2}\right ]^{1/2}
{M_\Lambda - M_n\over m_s -m_d}\left [ {m_s+m_d\over if_K M_K^2} \sqrt{2}
<0|H^\dagger_{eff}(Q_{11})|K^0>\right ]_{fig. 2.b}\;,\nonumber\\
\\
S(\Xi^0 \rightarrow \Lambda\pi^0) &=& -{i\over \sqrt{2}f_\pi} <\Lambda
|H^\dagger_{eff}(Q_{11})|\Xi^0>|_{fig.2.a} \nonumber\\
&-& {i\over \sqrt{2}f_\pi}\left [{3\over 2}\right ]^{1/2}
{M_\Xi - M_\Lambda\over m_s -m_d}\left [ {m_s+m_d\over if_K M_K^2} \sqrt{2}
<0|H^\dagger_{eff}(Q_{11})|K^0>\right ]_{fig. 2.b}\;.
\end{eqnarray}
We use pole model to calculate the P-wave amplitudes. For consistency, one
should include both the baryon and Kaon poles\cite{dh2}. The "tadpole"
contribution
from Fig. 2.b in the S-wave amplitude and the Kaon pole contribution in the
P-wave amplitude are both sizable. However, the CP violating observable
A depends on the difference between the S-wave phase $\phi^S$ and
the P-wave phase $\phi^P$. There is substantial cancellation between
the "tadpole" contribution in the S-wave and the Kaon pole contribution
in the P-wave.

The contributions to CP violation in hyperon decays from the same operator
$Q_{11}$ have been studied in the Weinberg model\cite{dp,dhp}. We can use some
of the results from there. However in the Weinberg model the CP violating
parameter $\bar \epsilon$ in the $K^0 -\bar K^0$ is also generated by the
operator $Q_{11}$ through long distance $\pi$, $\eta$ and $\eta'$ pole
contributions, the
coefficient $C_{11}(Weinberg)$ of $Q_{11}$\cite{dh} is fixed.
In our case, the contribution to $\bar \epsilon$ from $Q_{11}$ is small. To
obtain the CP violating phases in the decay
amplitudes, we
can repeat the MIT bag model calculations in Ref.\cite{dhp}.
In fact we can obtain the phases by replacing the matrix element
$<\pi^0| -{G_F\over \sqrt{2}} Im(V^*_{ud}V_{us}C^*_{11}(Weinberg))
Q^\dagger_{11}|K^0)>$ in
 the Weinberg model by
the SM matrix element, $<\pi^0|-H^\dagger_{eff}(Q_{11})|K^0>$, and rescale the
phases. We define the rescaling factor R as
\begin{eqnarray}
R &=&  {<\pi^0| -H^\dagger_{eff}(Q_{11})|K^0>\over
<\pi^0| -{G_F\over \sqrt{2}} Im(V^*_{ud}V_{us}C^*_{11}(Weinberg))
Q^\dagger_{11}|K^0)>}\nonumber\\
&=&-{G_F\over \sqrt{2}}y_{11}Im(V_{td}V_{ts}^*) {g_s\over 32\pi^2}
{m_s\tilde A_{K\pi}\over 5.8 \times 10^{-11}\; GeV^2}\;.
\end{eqnarray}
Here we have used $<\pi^0| -{G_F\over \sqrt{2}}
Im(V^*_{ud}V_{us}C^*_{11}(Weinberg))
Q^\dagger_{11}|K^0)> = 5.8 \times 10^{-11}\;GeV^2$\cite{dhp}.
We will follow Weinberg to use $g_s \approx 4\pi/\sqrt{6}$\cite{ww}.
The CP violating phases in hyperon decays from $Q_{11}$ in the SM are now
obtained by multiplying R to the phases obtained in Ref.\cite{dhp} for the
Weinberg model.

In Table 2 we list the CP violating observables A for
$\Lambda$ and $\Xi$ decays from different operators.
{}From Table 2, we see that the $Q_{11}$ contribution to CP violation in
$\Lambda \rightarrow N\pi$ is negligible. However the contribution to CP
violation in $\Xi \rightarrow \Lambda \pi$ can be important. The result is to
enhance CP violation in $\Xi$ decays by about 17\%. If we use the upper
value of $0.17 \;GeV^4$ for $m_s\tilde A_{K\pi}$, the
enhancement will be 25\%.  The allowed regions for $A(\Lambda)$ and
$A(\Xi)$ without the contributions from $Q_{11}$ are $3\times 10^{-5}\sim
0.4\times 10^{-5}$ and $5.3\times 10^{-5} \sim 0.7\times 10^{-5}$. With the new
contributions from $Q_{11}$ the allowed region for $A(\Lambda)$ is not changed,
but
the allowed region for $A(\Xi)$ becomes to $6.6\times 10^{-5} \sim 0.7 \times
10^{-5}$. This may be tested in the future\cite{exp}.

\acknowledgments
This work was supported in part by the Department of Energy Grant No.
DE-FG06-85ER40224 and DE-AN03-76SF00235. We thank E. Golowich for reading the
manuscript.

\begin{table}
\caption{$\epsilon'/\epsilon \times 10^4$ for $Im(V_{td}V_{ts}^*) =
2.77\times 10^{-4}(m^2_t/m_W^2)^{-0.365}$.}
\begin{tabular}{|c|c|c|c|c|}
$m_t$(GeV) & 130 &170 &200 &230\\ \hline
$\Lambda_4 = 200 \;MeV$   &&&&\\
$Q_1 - Q_{10}$& 6.3&2.8&0.7&-1.4\\
$Q_{11}$&0.65&0.58&0.5&0.48\\ \hline
$\Lambda_4 = 300\;MeV$&&&&\\
$Q_1-Q_{10}$&7.8&3.6&1.0&-1.7\\
$Q_{11}$&0.73&0.63&0.55&0.50
\end{tabular}
\label{table1}
\end{table}

\begin{table}
\caption{$A\times 10^{5}$ for $Im(V_{td}V_{ts}^*) =
2.77\times 10^{-4}(m^2_t/m_W^2)^{-0.365}$ and $m_s\tilde A_{K\pi}
= 0.12\;GeV^4$.}
\begin{tabular}{|c|c|c|c|c|}
$m_t$(GeV) & 130 &170 &200 &230\\ \hline
$\Lambda_4 = 200 \;MeV$   &&&&\\
$A(\Lambda^0_-) (Q_1 - Q_{10})$& -1.5&-1.3&-1.1&-1.0\\
$A(\Lambda^0_-)(Q_{11})$&-0.043&-0.037&-0.033&-0.031\\
$A(\Xi^-_-)(Q_1-Q_{10})$&-2.6&-2.2&-1.9&-1.8\\
$A(\Xi^-_-)(Q_{11})$&-0.55&-0.47&-0.41&-0.40\\ \hline
$\Lambda_4 = 300\;MeV$&&&&\\
$A(\Lambda^0_-) (Q_1 - Q_{10})$& -2.0&-1.7&-1.5&-1.4\\
$A(\Lambda^0_-)(Q_{11})$&-0.046&-0.039&-0.034&-0.032\\
$A(\Xi^-_-)(Q_1-Q_{10})$&-3.4&-2.9&-2.5&-2.3\\
$A(\Xi^-_-)(Q_{11})$&-0.58&-0.50&-0.44&-0.40\\
\end{tabular}
\label{table2}
\end{table}

\begin{picture}(100,200)(10,20)
\put(30, 100){\line(1,0){60}}
\put( 105, 105){\line(1,1){50}}
\put( 105, 95){\line(1,-1){50}}
\put( 90, 95){\framebox(15, 10){W}}
\put( 90, 10){\makebox(0,0){1.a}}
\put(40, 110){\makebox(0,0){K}}
\put(140, 50){\makebox(0,0){$\pi$}}
\put(140, 150){\makebox(0,0){$\pi$}}

\put(230, 100){\line(1,0){60}}
\put( 305, 105){\line(1,1){50}}
\put( 305, 95){\line(1,-1){50}}
\put( 290, 95){\framebox(15, 10){St}}
\put(305, 100){\line(1,0){40}}
\put( 345, 95){\framebox(15,10){W}}
\put( 290, 10){\makebox(0,0){1.b}}
\put(240, 110){\makebox(0,0){K}}
\put(335, 110){\makebox(0,0){K}}
\put(340, 50){\makebox(0,0){$\pi$}}
\put(340, 150){\makebox(0,0){$\pi$}}
\put(200, -20){\makebox(0,0){Fig. 1. Contributions to $K\rightarrow \pi\pi$
amplitude.  Here W and}}
\put(190, -40){\makebox(0,0){St indicate weak and strong interactions,
respectively.}}

\put(30, -150){\line(1,0){60}}
\put( 105, -145){\line(1,1){50}}
\put( 105, -155){\line(1,-1){50}}
\put( 90, -155){\framebox(15, 10){W}}
\put( 90, -240){\makebox(0,0){2.a}}
\put(40, -140){\makebox(0,0){$B_i$}}
\put(140, -200){\makebox(0,0){$\pi$}}
\put(140, -100){\makebox(0,0){$B_f$}}

\put(230, -150){\line(1,0){60}}
\put( 305, -145){\line(1,1){50}}
\put( 305, -155){\line(1,-1){50}}
\put( 290, -155){\framebox(15, 10){St}}
\put(305, -150){\line(1,0){40}}
\put( 345, -155){\framebox(15,10){W}}
\put( 290, -240){\makebox(0,0){2.b}}
\put(240, -140){\makebox(0,0){$B_i$}}
\put(335, -140){\makebox(0,0){K}}
\put(340, -200){\makebox(0,0){$\pi$}}
\put(340, -100){\makebox(0,0){$B_f$}}
\put(200, -270){\makebox(0,0){Fig. 2. Contributions to the S-wave
$B_i\rightarrow B_f\pi$ amplitudes.}}
\end{picture}

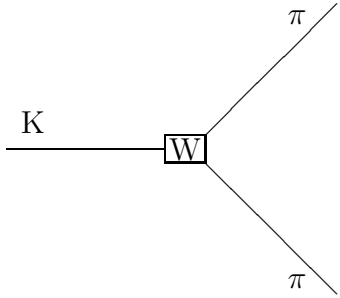
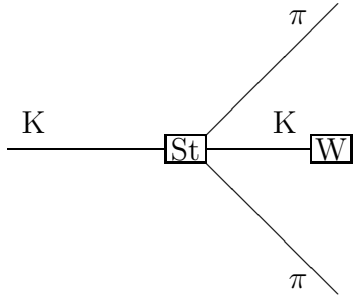
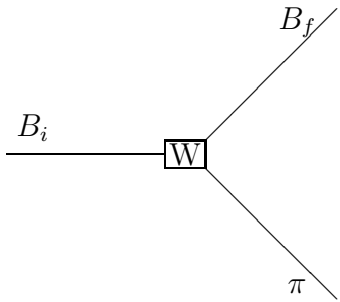
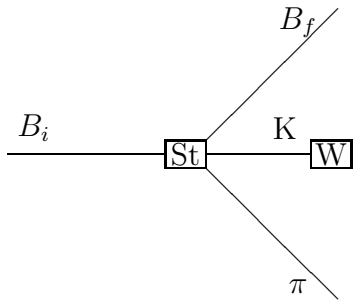
\end{document}